\documentclass[10pt,twocolumn,twoside]{IEEEtran} 
\usepackage{graphicx}
\usepackage{amsmath}
\usepackage{amssymb}
\usepackage{hyperref}
\usepackage{float}
\usepackage{setspace}
\usepackage{ragged2e}
\usepackage{tocloft}
\usepackage{subcaption}
\usepackage{mathbbol}
\usepackage{xcolor}
\usepackage{pifont} 
\usepackage{amsfonts}

\usepackage{amsthm}
\newtheorem{theorem}{Theorem}
\newtheorem{lemma}{Lemma}
\newtheorem{remark}{Remark}
\newtheorem{proposition}{Proposition}

\ifCLASSOPTIONcompsoc
  \usepackage[nocompress]{cite}
\else
  \usepackage{cite}
\fi
\def\BibTeX{{\rm B\kern-.05em{\sc i\kern-.025em b}\kern-.08em
    T\kern-.1667em\lower.7ex\hbox{E}\kern-.125emX}}
\begin{document}
\title{Adaptive Collective Decision-Making in Signed Networks}
%
%
%
%


\author{Vishnudatta Thota,
        Aditi Saxena,
        Twinkle Tripathy*, and 
        Vaibhav Srivastava
\IEEEcompsocitemizethanks{\IEEEcompsocthanksitem V. Thota, A. Saxena, and T. Tripathy are with the Department of Electrical Engineering, Indian Institute of Technology Kanpur, Uttar Pradesh, India, 208016.\protect\\
*Correspondence and requests for materials should be addressed to T. Tripathy (email: ttripathy@iitk.ac.in)
\IEEEcompsocthanksitem V. Srivastava is with the Electrical and Computer Engineering, Michigan State University, East Lansing, MI, USA, 48834.}
}



\IEEEtitleabstractindextext{%
\begin{abstract}
\justifying   
Collective decision-making often unfolds in networks where cooperation and antagonism coexist. Such signed interactions arise in social, biological, and robotic systems, and can lead to qualitatively different collective outcomes: consensus, polarization, or clustering. Existing signed network models typically explain these outcomes separately, leaving open how they can emerge within a single dynamical framework. Here, we introduce a generalized Laplacian flow for bi-partitioned signed networks, a class of networks that allows cooperative and antagonistic interactions both within and across two prescribed groups. The proposed model includes an adaptive social effort state, interpreted as a leaky memory of persistent signed network disagreement, which enables bounded and nontrivial steady-states beyond classical signed Laplacian flows. We derive conditions that determine whether the network reaches consensus, polarization, or clustering based on the spectral structure of the generalized Laplacian and effective resistance matrices. The framework applies beyond structurally balanced, weakly balanced, and quasi-structurally balanced networks, and reveals how changes in signed interaction strengths can shift a network between collective decision regimes. We validate the theory on stochastic block model networks, a judicial voting network, and a U.S. senate bill co-sponsorship network. Altogether, our results provide a structural and dynamical account of how mixed cooperative and antagonistic relationships shape emergent collective behaviors.
\end{abstract}
}

\maketitle
\IEEEdisplaynontitleabstractindextext
\IEEEpeerreviewmaketitle

{\section{Introduction}\label{sec:introduction}}
{C}{ollective} decision-making in \textcolor{black}{the expansive area of opinion dynamics} describes how individuals with partial information can, through interaction, produce coherent group-level outcomes. Through repeated social, biological, or engineered interactions, agents revise their opinions, preferences, or actions and may eventually reach consensus, separate into opposing camps, or form multiple clusters. In many such systems, influence is not purely cooperative. Individuals may support, trust, or align with some neighbors while opposing, distrusting, or repelling others. Signed networks provide a natural representation of this co-existence of agreement and disagreement, with positive edges denoting cooperative ties and negative edges denoting antagonistic ties.

The study of signed networks has its roots in social psychology and graph theory. Structural balance theory was introduced to formalize social configurations consistent with Heider's balance principles \cite{cartwright1956structural,harary1953notion,heider1958psychology}, wherein antagonism is organized into a coherent two-camp structure. Mathematically, a strongly balanced signed network admits a bipartition of the node set into two disjoint groups such that \textcolor{black}{the} interactions within each group are positive and \textcolor{black}{that} across groups are negative. This structure naturally supports polarization into two opposing camps.
However, strongly balanced networks represent only a restricted class of signed networks. Weak balance relaxes this structure and allows the network to split into multiple mutually antagonistic clusters \cite{Clustering1967}. Empirical studies further show that real social networks may be neither strongly nor weakly balanced \cite{estrada2014social,facchetti2011computing}. 
\textcolor{black}{A recent work on quasi-structurally balanced networks \cite{shi2023opinion} studies polarization in networks with two antagonistic subsets and mixed intra-subset interactions. In such networks, an antagonistic interaction between two nodes within the same subset is permitted only if a positive path exists between them. This assumption on network topology is further relaxed in \cite{thota2025dominant}, which allows arbitrary antagonistic interactions within each subset and analyzes asymmetric polarization in the presence of a dominant group.}

A central question in the area of opinion dynamics is, therefore, how the signed structure of a network \textcolor{black}{shapes the steady-state opinion formation}. Classical models often address this question through signed Laplacian flows. Unlike unsigned networks, signed networks admit multiple Laplacian definitions, two of the most common being the opposing and repelling Laplacians \cite{shi2019dynamics}. 
\textcolor{black}{The opposing Laplacian explains bipartite consensus, or polarization in structurally balanced networks} \cite{altafini2012consensus}. The opposing-Laplacian-based models have also been used to study controlled clustering in arbitrary signed networks \cite{priya2024desired}. The repelling Laplacian, by contrast, models negative interactions as repulsive couplings. Its flow can converge to consensus when antagonistic interactions are sufficiently \textcolor{black}{weak relative to the cooperative ones} \cite{fontan2022multiagent}. \textcolor{black}{Thus, opposing and repelling Laplacian flows capture contrasting effects of signed interactions,} but they do not by themselves provide a unified account of consensus, polarization, and clustering in signed networks.

This limitation suggests that collective outcomes are not determined solely by the polarity of edge-weights. The same pattern of cooperation and antagonism may lead to different final states depending on how individuals respond to peer interactions and how they regulate their own opinions. \textcolor{black}{In nonlinear models of collective decision-making, the latter is modelled as \emph{social effort} term which explains how individual behaviours respond to the social environment \cite{fontan2018achieving}. In structurally unbalanced networks, the inclusion of this term results in more realistic behaviours such as polarisation and clustering rather than divergence and trivial consensus \cite{fontan2021role}.}  

Motivated by the above-mentioned works, in this paper, we study opinion dynamics on \emph{bi-partitioned signed networks} (BSNs). A BSN consists of a prescribed partition of the node set into two disjoint subsets, while allowing both cooperative and antagonistic interactions within and across the subsets. This class includes structurally balanced networks~\cite{altafini2012consensus}, quasi-structurally balanced networks~\cite{shi2023opinion}, and generalized quasi-structurally balanced networks~\cite{thota2025dominant} as special cases, while allowing more general signed interaction patterns. BSNs arise naturally in systems with two broad groups, communities, or roles, but where interaction polarities do not follow an idealized balance pattern. Examples include two-party political systems, social networks with two broad communities, and bipartite interaction systems in recommendation, biological, or socio-technical networks~\cite{fraxanet2024unpacking,tan2026compensation,gaspar2019drugtarget,harbridge2023bipartisan}. In such systems, individuals may cooperate across groups, disagree within groups, or exhibit mixed patterns of support and antagonism~\cite{harbridge2023bipartisan, andris2015rise,elinson2023intraparty}. Depending on the relative structure and the strength of these interactions, real world networks may converge to consensus, polarize across the prescribed partition, or form multiple opinion clusters.

The main contribution of this paper is to introduce and analyze a generalized Laplacian flow for arbitrary BSNs. Within this framework, consensus, polarization, and clustering arise as distinct steady-state behaviors of the same dynamical model. We derive spectral and effective resistance-based conditions that determine which outcome emerges from the signed interaction structure. Finally, \textcolor{black}{we show that slight changes in the magnitudes and polarities of edge weights (or interactions)} can shift the collective behavior among consensus, polarization, and clustering.

The rest of the paper is organized as follows. In Section~\ref{sec:Preliminaries}, we review preliminaries regarding signed networks. Subsequently, in Section~\ref{sec:Bi-partitioned signed networks}, we define BSNs\footnote{Structurally balanced networks are a special case of BSNs in which all intra-subset edges are cooperative and all inter-subset edges are antagonistic \cite{heider1958psychology,altafini2012consensus}.}. Sections~\ref{sec:Generalized Laplacian flow for collective decision dynamics} and~\ref{sec:Collective decision outcomes in bi-partitioned signed networks} discuss the generalized Laplacian flow and its collective decision outcomes, respectively. \textcolor{black}{The structural conditions for achieving transitions in collective decision outcomes are discussed in Section~\ref{sec:Structural transitions between collective outcomes}}. Thereafter, we validate the theory using simulations on synthetic and empirical signed networks in Section~\ref{sec:Numerical and empirical validation}. Section~\ref{sec:Discussion and Future Works} concludes the paper and discusses directions for future research.
\section{Preliminaries}
\label{sec:Preliminaries}
\textit{Notations: }Let $\mathbb{R}$, $\mathbb{R}^{+}$, and $\mathbb{R}^n$ denote the set of real numbers, the set of positive real numbers, and the $n$-dimensional real vector space, respectively. The vector $\mathbf{1}_n\in\mathbb{R}^n$ denotes the vector of all ones, and $I_n$ denotes the $n\times n$ identity matrix. The spectrum of a matrix $M$ is denoted by $\sigma(M)$. For a real symmetric matrix $M$, we write $M\succ 0$ if $M$ is positive definite and $M\succeq 0$ if $M$ is positive semidefinite.

A signed weighted network is denoted by $\mathcal G=(\mathcal V,\mathcal E,A)$, where $\mathcal V=\{1,\ldots,n\}$ is the node set, $\mathcal E\subseteq \mathcal V\times \mathcal V$ is the edge set, and $A=[a_{ij}]\in\mathbb R^{n\times n}$ is the signed adjacency matrix. A positive edge weight $a_{ij}>0$ represents a cooperative interaction, a negative edge weight $a_{ij}<0$ represents an antagonistic interaction, and $a_{ij}=0$ indicates the absence of an edge. Throughout this paper, unless otherwise stated, $\mathcal G$ is undirected, so that $A=A^\top$.

Let $\mathcal G_+=(\mathcal V,\mathcal E_+)$ and $\mathcal G_-=(\mathcal V,\mathcal E_-)$ denote the positive-edge and negative-edge subgraphs of $\mathcal G$, respectively. For an undirected network, a spanning forest is a subgraph that contains all nodes and consists of a spanning tree for each connected component. Let $\mathcal F_-$ be a spanning forest of $\mathcal G_-$, and let $\mathcal C_-$ denote the subgraph consisting of the remaining negative edges. Then the signed network can be decomposed as $\mathcal G=\mathcal F_- \cup \mathcal C_- \cup \mathcal G_+$. Further, we assign an arbitrary orientation to each edge and let $B\in\mathbb R^{n\times m}$ be the corresponding incidence matrix. With a suitable ordering of the edges, write $B=
\begin{bmatrix}
B_{\mathcal F_-} & B_{\mathcal C_-} & B_{\mathcal G_+}
\end{bmatrix}$, 
where the three blocks correspond to edges in $\mathcal F_-$, $\mathcal C_-$, and $\mathcal G_+$, respectively.

\textcolor{black}{We use the following terminology for limiting opinion profiles, where the vector of opinions is denoted by $\mathbf{x}=[x_1,x_2,\ldots,x_n]^T$. Opinions converge to consensus if $x_i(t)-x_j(t)\to 0$ for all $i,j\in\mathcal V$. They converge to polarization with respect to a bipartition $\mathcal V=\mathcal V_1\cup\mathcal V_2$ if $x_i(t)-x_j(t)\to 0$ for all $i,j$ in the same subset and $x_i(t)+x_j(t)\to 0$ for all $i\in\mathcal V_1$, $j\in\mathcal V_2$. Further, clustering occurs if the limiting opinions form more than one cluster but do not satisfy the polarization condition for two groups.}
\vspace{-2mm}
\section{Bi-partitioned signed networks}\label{sec:Bi-partitioned signed networks}
Consider a group of individuals interacting over a connected network, represented by
\(\mathcal G=(\mathcal V,\mathcal E,A)\). In the given framework, the network is signed and is equipped with a prescribed bipartition of the node set $\mathcal{V}$ into two disjoint subsets \(\mathcal V_1\) and \(\mathcal V_2\), with \(\mathcal V=\mathcal V_1\cup\mathcal V_2\) and \(\mathcal V_1\cap\mathcal V_2=\emptyset\). We refer such networks as BSNs. These networks allow both cooperative and antagonistic interactions within and across the two subsets. \textcolor{black}{ Our objective is to propose a Laplacian-based model that explains the commonly observed behaviors of consensus, polarization, and clustering in BSNs.} 

The bipartitioning in BSNs generally arise from external information, such as party labels, ideological groups, biological classes, or user--item types. In this work, \textcolor{black}{we consider such BSNs, where the bipartition is given as a part of the problem data}. For each individual \(i\in\mathcal V_k\), \(k\in\{1,2\}\), we distinguish its neighbors according to whether they lie in the same subset or the other {one}, and whether the corresponding interaction is cooperative or antagonistic. We denote the intra-subset cooperative and antagonistic neighbor sets by \(\mathcal N_i^+\) and \(\mathcal N_i^-\), respectively. Similarly, we denote the inter-subset cooperative and antagonistic neighbor sets by \(\hat{\mathcal N}_i^+\) and \(\hat{\mathcal N}_i^-\), respectively. The complete neighbor set is $\mathcal N_i
=
\mathcal N_i^+
\cup
\mathcal N_i^-
\cup
\hat{\mathcal N}_i^+
\cup
\hat{\mathcal N}_i^-$.
For a node $i\in\mathcal V_r$, $r\in\{1,2\}$, the precise set-theoretic definitions of these neighbor classes are defined as follows:
\begin{align*}
 \mathcal N_i^+
&=
\{j:(i,j)\in\mathcal E,\; j\in\mathcal V_r,\; a_{ij}>0\},  \\ \mathcal N_i^-
&=
\{j:(i,j)\in\mathcal E,\; j\in\mathcal V_r,\; a_{ij}<0\}, \\
\hat{\mathcal N}_i^+
&=
\{j:(i,j)\in\mathcal E,\; j\notin\mathcal V_r,\; a_{ij}>0\}, \\  \hat{\mathcal N}_i^-
&=
\{j:(i,j)\in\mathcal E,\; j\notin\mathcal V_r,\; a_{ij}<0\}.
\end{align*}

Using these four neighbor classes, we define two node-level signed connectivity indices. The inter-subset connectivity index \(\alpha_i\) is the net signed influence on node \(i\) from the opposite subset, and the intra-subset connectivity index \(\beta_i\) is the net signed influence on node \(i\) from its own subset:
\[
\alpha_i
=
\sum_{j\in\hat{\mathcal N}_{i}^{+}\cup\hat{\mathcal N}_{i}^{-}} a_{ij},
\quad \text{and} \quad 
\beta_i
=
\sum_{j\in\mathcal N_{i}^{+}\cup\mathcal N_{i}^{-}} a_{ij}.
\]
\textcolor{black}{In a BSN, these} indices summarize whether cooperation or antagonism dominates within or across a prescribed bipartition.

A BSN is called type-1 if \(\alpha_i\ge0\) and \(\beta_i\ge0\) for all \(i\in\mathcal V\). In this case, both inter-subset and intra-subset net influences are cooperative. \textcolor{black}{It} is type-2 if \(\alpha_i<0\) and \(\beta_i\ge0\) for all \(i\in\mathcal V\). In this case, the inter-subset interactions are net antagonistic, while the intra-subset \textcolor{black}{ones} are net cooperative. All the remaining BSNs are type-3. This classification is relative to the prescribed bipartition. As shown subsequently in the manuscript, it helps determine whether the proposed dynamics (in Section \ref{sec:Generalized Laplacian flow for collective decision dynamics}) produces consensus, polarization, or clustering at steady-state.
\section{Generalized Laplacian flow for collective decision dynamics}
\label{sec:Generalized Laplacian flow for collective decision dynamics}
Consider a connected BSN with $n$ individuals partitioned into
$\mathcal{V}_{1}=\{1,\ldots,q\}$ and
$\mathcal{V}_{2}=\{q+1,\ldots,n\}$, where
$|\mathcal{V}_{1}|=q$ and $|\mathcal{V}_{2}|=n-q$. Let
$\mathbf{x}=[x_1,x_2,\ldots,x_n]^T\in\mathbb{R}^{n}$ denote the vector of individual opinions on a given issue. The opinion of each individual evolves through its signed interactions with neighboring individuals.

Classical signed Laplacian flows provide important limiting models for signed-network opinion dynamics, but they do not capture consensus, polarization, and clustering within a single bounded dynamical framework. We therefore introduce an adaptive social effort state, denoted by \(\mathbf u\in\mathbb R^n\), which retains a leaky memory of the generalized disagreement induced by the signed network.
This memory-like state prevents the opinion dynamics from being governed only by instantaneous signed interactions and allows the system to reach bounded, nontrivial collective outcomes.

We first define the generalized degree matrix $D_g$ by,

\begin{equation}
\label{eq:ouut_degree_BSN}
[D_g]_{ii}
=
\left|
\sum_{j\in\mathcal{N}_{i}^{+}\cup\mathcal{N}_{i}^{-}} a_{ij}
\right|
+
\left|
\sum_{j\in\hat{\mathcal{N}}_{i}^{+}\cup\hat{\mathcal{N}}_{i}^{-}} a_{ij}
\right|.
\end{equation}
The generalized Laplacian matrix is then defined as $L_g=D_g-A$.
This matrix incorporates the net intra-subset and inter-subset signed influences that determine the collective decision outcome.

We model the opinion and social effort dynamics using the generalized Laplacian flow: $\dot{\mathbf{x}}=-L_g\mathbf{x}+\mathbf{u}$, $\tau\dot{\mathbf{u}}=L_g\mathbf{x}-\mathbf{u}$,
where $\mathbf{u}(0)=\mathbb{0}_n$ and $\tau>0$ is the time constant of the social effort dynamics. \textcolor{black}{The adaptive social effort preserves the information in the initial condition while preventing the steady-state behavior from diverging or from converging to the neutral state.} Equivalently, we have;
\begin{equation}\label{eq:opinion_update_rule}
\begin{bmatrix}
\dot{\mathbf{x}} \\
\dot{\mathbf{u}}
\end{bmatrix} =
\begin{bmatrix}
- L_{g} & I \\
\frac{1}{\tau} L_{g} & -\frac{1}{\tau} I
\end{bmatrix}
\begin{bmatrix}
\mathbf{x} \\
\mathbf{u}
\end{bmatrix}.
\end{equation}
Here, \(\mathbf u\) is the adaptive social effort, which acts as a leaky integrator of the generalized disagreement \(L_g\mathbf x\). 
In the existing literature, social effort is used to describe the strength of an individual's commitment or attention to a social interaction \cite{fontan2021role, fontan2018achieving, gray2018multiagent, leonard2024fast}. In the present model, the \(i\)th entry of \(\mathbf u\) represents the adaptive effort exerted by individual \(i\) in response to the disagreement experienced by that individual in signed networks. 

\textcolor{black}{We characterize the spectral consequences of this augmented flow to understand how its eigenvalue structure influences the steady-state behavior and stability of the system in the following section.} 
\subsection{Stability of the Generalized Laplacian Flow}
\begin{lemma}[Generalized Laplacian flow]
\label{lem:augmented_flow_properties}
\textcolor{black}{Consider a connected undirected BSN following the generalized Laplacian flow~\eqref{eq:opinion_update_rule} with $\mathbf u(0)=0$ and generalized Laplacian matrix $L_{g}$.}
Let $L_g v_i=\lambda_i v_i$, where the eigenvectors $v_i, ~ i \in \{1,\cdots,n\}$ form an orthonormal basis. Then the following statements hold:
\begin{enumerate}
    \item For each eigenvalue $\lambda_i$ of $L_g$, the corresponding two-dimensional modal dynamics has the following characteristic polynomial,
    \[
    p_i(s)
    =
    s\left(s+\lambda_i+\frac{1}{\tau}\right).
    \]
    Hence each mode has one zero eigenvalue and one eigenvalue $s_i=-\left(\lambda_i+\frac{1}{\tau}\right)$. 
    The nonzero modal eigenvalue is Hurwitz if and only if $(1+\tau\lambda_i)>0$. Consequently, all nonzero modal dynamics are stable if and only if $ (I_n+\tau L_g)\succ0$. 

    \item Along the generalized Laplacian flow, $\mathbf x(t)+\tau\mathbf u(t)=\mathbf x(0)$, $t\ge0$.

    \item If $(I_n+\tau L_g)\succ0$, then every trajectory converges to an equilibrium, and the limiting opinion vector is $\mathbf x_\infty=(I_n+\tau L_g)^{-1}\mathbf x(0)$.
    Equivalently, if $\mathbf x(0)=\sum_{i=1}^n c_i v_i$, then $\mathbf x_\infty
    =
    \sum_{i=1}^n
    \frac{c_i}{1+\tau\lambda_i}v_i$.
\end{enumerate}
\end{lemma}
\begin{proof}
\textcolor{black}{Since the BSN is undirected with real interaction weights, the corresponding $L_{g}$ is real symmetric, and its eigenvectors form an orthonormal basis.}
Projecting $\mathbf x$ and $\mathbf u$ onto an eigenvector $v_i$ gives,
\[
\begin{bmatrix}
\dot x_i\\
\dot u_i
\end{bmatrix}
=
\begin{bmatrix}
-\lambda_i & 1\\
\tau^{-1}\lambda_i & -\tau^{-1}
\end{bmatrix}
\begin{bmatrix}
x_i\\
u_i
\end{bmatrix}.
\]
The characteristic polynomial of this modal matrix is,
\[
p_i(s)
=
\det
\begin{bmatrix}
s+\lambda_i & -1\\
-\tau^{-1}\lambda_i & s+\tau^{-1}
\end{bmatrix}
=
(s+\lambda_i)(s+\tau^{-1})-\tau^{-1}\lambda_i.
\]
Therefore, $p_i(s)
=
s^2+s\left(\lambda_i+\frac{1}{\tau}\right)
=
s\left(s+\lambda_i+\frac{1}{\tau}\right)$.

Thus each modal system has a zero eigenvalue and a nonzero eigenvalue $s_i=-(\lambda_i+\tau^{-1})$. The nonzero eigenvalue is Hurwitz if and only if $(\lambda_i+\tau^{-1})>0$, equivalently $(1+\tau\lambda_i)>0$. 
This condition holds for all $i$ if and only if $(I_n+\tau L_g)\succ0$.

Next, using the augmented dynamics,
\[
\dot{\mathbf x}+\tau\dot{\mathbf u}
=
(-L_g\mathbf x+\mathbf u)
+
\tau(\tau^{-1}L_g\mathbf x-\tau^{-1}\mathbf u)
=0.
\]
Thus, $\mathbf x(t)+\tau\mathbf u(t)$ is conserved. Since $\mathbf u(0)=0$, it follows that $\mathbf x(t)+\tau\mathbf u(t)=\mathbf x(0)$, $t\ge0$.

Finally, suppose $(I_n+\tau L_g)\succ0$. Then, all the nonzero eigenvalues are Hurwitz, so every trajectory converges to the equilibrium manifold. At equilibrium, $0=-L_g\mathbf x_\infty+\mathbf u_\infty$, 
and hence $\mathbf u_\infty=L_g\mathbf x_\infty$. Combining this identity with the conserved quantity gives $\mathbf x_\infty=(I_n+\tau L_g)^{-1}\mathbf x(0)$.
The modal expression follows by expanding $\mathbf x(0)$ in the eigenbasis of $L_g$.
\end{proof}
\begin{remark}
The generalized Laplacian flow is closely related to the model introduced in \cite{thota2025dominant} for generalized quasi-structurally balanced networks with a dominant group. \textcolor{black}{However, the present formulation extends this framework to arbitrary BSNs by defining the generalized degree in terms of the intra-subset and inter-subset net connectivity indices.}
\end{remark}
\textcolor{black}{Now, we provide the relationship between our proposed Laplacian flows with the existing ones in the next section.
This relationship will help the readers understand the theoretical relevance of the proposed formulation and clarify its connection to existing models.}
\subsection{Relationship with the repelling and opposing Laplacian flows}
\label{sec:Relationship with repelling and opposing Laplacian flows}
The repelling signed Laplacian \cite{fontan2022multiagent} and the opposing signed Laplacian \cite{altafini2012consensus,shi2019dynamics} are defined,
\begin{align*}
[L_r]_{ij}
&=
\begin{cases}
-a_{ij}, & i\neq j,\\
\sum_{j\in\mathcal N_i}a_{ij}, & i=j,
\end{cases}
\quad \text{and} \\  
[L_o]_{ij}
&=
\begin{cases}
-a_{ij}, & i\neq j,\\
\sum_{j\in\mathcal N_i}|a_{ij}|, & i=j,
\end{cases}
\end{align*}
where $\mathcal N_i$ denotes the set of neighbors of node $i$. The associated repelling and opposing Laplacian flows,
\[
\dot{\mathbf x}=-L_r\mathbf x
\quad \text{and} \quad
\dot{\mathbf x}=-L_o\mathbf x, ~~\text{respectively.}
\]

\textcolor{black}{These two flows correspond to two distinct models of negative interaction. For structurally balanced signed networks, the opposing flow produces bipartite consensus; for structurally unbalanced networks, the limiting state is the neutral opinion \cite{altafini2012consensus,shi2019dynamics}. In the repelling flow, a negative edge acts as a repulsive interaction relative to the neighboring state. When the positive subgraph is sufficiently connected and the negative interaction strength is small, the flow can still converge to consensus, while stronger negative interactions can produce divergence \cite{shi2019dynamics,fontan2022multiagent}.}

\textcolor{black}{The following result presents the conditions under which the repelling Laplacian has the spectral structure required for achieving consensus.}

\begin{lemma}[Characterization of effective resistance\cite{chen2016characterizing_effective_resistance}]
\label{lem:effective_resistance}
Let $L_r$ be the repelling signed Laplacian of a connected undirected signed network $\mathcal G$. Then $L_r\succeq 0$ with a simple zero eigenvalue if and only if the effective resistance matrix,
\[
\Gamma_{\mathcal F_-}
=
B_{\mathcal F_-}^{\top}L_r^\dagger B_{\mathcal F_-}
\]
is positive definite, where $L_r^\dagger$ denotes the Moore--Penrose pseudoinverse of $L_r$.
\end{lemma}

Thus, the opposing and repelling flows capture different signed-interaction mechanisms, but neither directly provides a model that distinguishes consensus, polarization, and clustering for arbitrary networks.

We next relate the proposed generalized Laplacian to these classical flows. 
\begin{itemize}
\item For a type-1 BSN, $\alpha_i\ge0$ and $\beta_i\ge0$. Therefore,
\[
[D_g]_{ii}
=
\alpha_i+\beta_i
=
\sum_{j\in\mathcal N_i}a_{ij},
\]
and, hence, $L_g=L_r$. Thus, when the social effort state is absent or frozen
(for example, in the singular limit $\tau\to\infty$ with $\mathbf u(0)=0$), the $\mathbf x$-dynamics reduce to the repelling
Laplacian flow.
\item For a type-2 BSN, an analogous statement holds after applying the sign transformation associated with the prescribed bipartition \cite{altafini2012consensus}. Order nodes so that $\mathcal V_1=\{1,\ldots,q\}$ and $\mathcal V_2=\{q+1,\ldots,n\}$, and define $R=\operatorname{diag}(-\mathbf 1_q,\mathbf 1_{n-q})$, $\mathbf z=R\mathbf x$, and $A_z=RAR^{-1}$. 
Let $L_z=D_g-A_z$ denote the transformed generalized Laplacian. The transformation reverses the polarities of inter-subset interactions and preserves the polarities of the intra-subset interactions. Hence, the transformed type-2 BSN \(\mathcal G_z\) has nonnegative net inter-subset and intra-subset connectivity. Consequently, $L_z$ coincides with the repelling Laplacian of the transformed signed network. 
\textcolor{black}{Thus, consensus of the opinions in the $\mathbf z$-domain corresponds to polarization of opinions in the $\mathbf x$-domain.}
\end{itemize}
\textcolor{black}{This section discussed the stability conditions of the proposed generalized Laplacian flows and their relation to the existing repelling and opposing Laplacian flows. Next, we derive the conditions under which different collective decision outcomes, such as consensus, polarization, and clustering, emerge.}
\section{Collective decision outcomes in bi-partitioned signed networks}
\label{sec:Collective decision outcomes in bi-partitioned signed networks}
The collective decision outcomes in BSNs are determined by the spectral structures induced by the generalized Laplacian \(L_g\) and the time constant $\tau$ of the social effort dynamics. Since the network is undirected, \(L_g\) is symmetric, and its eigenvalues are real. As shown in Lemma~\ref{lem:augmented_flow_properties}, the augmented dynamics in \eqref{eq:opinion_update_rule} are stable in all nonzero modes whenever $(I+\tau L_g)\succ0$, or equivalently \(1+\tau\lambda_i(L_g)>0\) for every eigenvalue \(\lambda_i(L_g)\). When \(L_g\) is not positive semidefinite, the dynamics can be made stable by choosing \(\tau\) sufficiently small. For \(\mathbf u(0)=\mathbb 0_n\), the resulting bounded equilibrium is,
\[
\mathbf{x}_f=(I+\tau L_g)^{-1}\mathbf{x}_0,
\]
where \(\mathbf{x}_0\) is the initial opinion vector.
Equivalently, each eigenmode of \(L_g\) with eigenvalue \(\lambda_i\) is weighted at equilibrium by the factor \((1+\tau\lambda_i)^{-1}\). In regimes where the large values of $\tau$ are admissible, only the zero-eigenvalue modes persist, so the final opinion approaches the projection of \(\mathbf{x}_0\) onto the nullspace of \(L_g\). The dimension and structure of this nullspace distinguish consensus, polarization, and clustering.
%

Next, we state the main result of this article on the generalized Laplacian flows.
\begin{theorem}[Collective decision outcomes in BSNs]
\label{thm:main_convergence}
Consider a connected undirected BSN $(\mathcal G,\mathcal V_1,\mathcal V_2)$ with at least one inter-subset edge. Let the opinion dynamics be governed by the generalized Laplacian flow~\eqref{eq:opinion_update_rule}, and suppose $(I_n+\tau L_g)\succ0$. Then the following regimes hold.
\begin{enumerate}
    \item \emph{Consensus-compatible regime:}
    If $\mathcal G$ is type-1, then $L_g=L_r$. Therefore, the limiting profile converges to consensus as $\tau\to\infty$ if and only if the effective resistance matrix $\Gamma_{\mathcal F_-}$ of the spanning forest formed by the negative edges of $\mathcal G$ is positive definite. In that case,
    \[
    \mathbf x_\infty
    \to
    \frac{1}{n}\mathbf 1_n\mathbf 1_n^\top \mathbf x(0)
    \qquad \text{as } \tau\to\infty.
    \]
\item \emph{Polarization-compatible regime:}
If $\mathcal G$ is type-2, then the limiting profile converges to
polarization as $\tau\to\infty$ if and only if the effective resistance matrix
$\Gamma_{\mathcal F_{z-}}$ of the spanning forest formed by the negative edges
of the transformed network $\mathcal G_z$ is positive definite. In that case,
\[
\mathbf x_\infty
\to
\frac{1}{n}R^{-1}\mathbf 1_n\mathbf 1_n^\top R\mathbf x(0)
\qquad \text{as } \tau\to\infty.
\]
    \item \emph{Clustering-compatible regime:}
    If the relevant effective resistance condition in the type-1 or type-2 case fails, or if the BSN is type-3, then the limiting profile is not constrained to a single consensus or two-faction polarization subspace. For $\tau$ such that $(I_n+\tau L_g)\succ0$, the equilibrium remains $\mathbf x_\infty=(I_n+\tau L_g)^{-1}\mathbf x(0)$.
    When $L_g\succeq0$ has a nullspace of dimension greater than one, the large-$\tau$ limit is the projection of $\mathbf x(0)$ onto $\ker L_g$, yielding a clustered profile.
\end{enumerate}
\end{theorem}
\begin{proof}
Consider a type-1 BSN. Since $\alpha_i\ge0$ and $\beta_i\ge0$, we have $L_g=L_r$. By Lemma~\ref{lem:effective_resistance}, $L_g\succeq0$ with a simple zero eigenvalue if and only if $\Gamma_{\mathcal F_-}\succ0$. In this case, $\ker L_g=\operatorname{span}\{\mathbf 1_n\}$.
Writing $\mathbf x(0)=c_1\mathbf 1_n+\sum_{i=2}^n c_i v_i$, with $\lambda_i>0$ for $i\ge2$, Lemma~\ref{lem:augmented_flow_properties} gives $\mathbf x_\infty
=
c_1\mathbf 1_n+
\sum_{i=2}^n
\frac{c_i}{1+\tau\lambda_i}v_i$.
Taking $\tau\to\infty$ eliminates all positive modes and gives $\mathbf x_\infty
\to
\frac{1}{n}\mathbf 1_n\mathbf 1_n^\top\mathbf x(0)$.
Thus, the limiting opinion profile is consensus.

Next, consider a type-2 BSN. Recall the z-domain transformation $\mathbf z=R\mathbf x$ from Section~\ref{sec:Relationship with repelling and opposing Laplacian flows}. 
The transformation reverses the polarities of inter-subset edges and preserves the
polarities of intra-subset edges. Hence, for a type-2 BSN, the transformed dynamics
are consensus-compatible in the $\mathbf z$-coordinates. By applying
Lemma~\ref{lem:effective_resistance} to the transformed network, $L_z\succeq0$
with a simple zero eigenvalue holds if and only if
$\Gamma_{\mathcal F_{z-}}\succ0$. In this case, $\ker L_z=\operatorname{span}\{\mathbf 1_n\}$. 
Therefore,
\[
\mathbf z_\infty
\to
\frac{1}{n}\mathbf 1_n\mathbf 1_n^\top \mathbf z(0)
=
\frac{1}{n}\mathbf 1_n\mathbf 1_n^\top R\mathbf x(0)
\qquad \text{as } \tau\to\infty.
\]
Transforming back to the original coordinates gives $\mathbf x_\infty
=
R^{-1}\mathbf z_\infty
\to
\frac{1}{n}R^{-1}\mathbf 1_n\mathbf 1_n^\top R\mathbf x(0)$,
which is a polarized profile with respect to the prescribed bipartition.

Finally, suppose that the relevant effective resistance condition fails, or that the BSN is type-3. In these cases, the generalized Laplacian does not reduce to a consensus-compatible repelling Laplacian in either the original coordinates or the transformed coordinates. If $L_g\succeq0$ and $\dim\ker L_g=k>1$, then,
\[
\mathbf x_\infty
=
\operatorname{Proj}_{\ker L_g}\mathbf x(0)
+
\sum_{\lambda_i>0}
\frac{c_i}{1+\tau\lambda_i}v_i 
\]\[
\mathbf x_\infty
\to
\operatorname{Proj}_{\ker L_g}\mathbf x(0)
\qquad
\text{as } \tau\to\infty.
\]
where $\operatorname{Proj}_{\ker L_g}$ denotes projection onto $\ker L_g$.
The resulting limiting profile generally contains multiple opinion clusters.
\textcolor{black}{If $\ker L_g$ contains vectors beyond the consensus direction $\mathbf{1}_{n}$, then $\operatorname{Proj}_{\ker L_g}\mathbf x(0)$ may have a nonzero component orthogonal to $\mathbf{1}_{n}$. Such components lead to the opinion convergence to multiple clusters instead of the consensus.}

If $L_g$ has negative eigenvalues, then the admissibility of the finite-$\tau$ equilibrium requires $1+\tau\lambda_i>0$ for all $i$. 
Equivalently, if $\lambda_{\min}(L_g)<0$, then $\tau<-\frac{1}{\lambda_{\min}(L_g)}$.
When this condition fails, the convergence result does not apply, and the corresponding unregularised modes may diverge. Hence, proved.
\end{proof}
\begin{remark}
\textcolor{black}{Consensus and polarization can be achieved as $\tau \rightarrow \infty$, which corresponds to the absence of social effort in Equation~\ref{eq:opinion_update_rule}. On the other hand, when $\lambda_{\min}(L_g)<0$, $\tau$ is upper-bounded, indicating that a finite level of social effort is necessary to achieve clustering. Thus, social effort plays a crucial role in achieving clustering, and its absence may result in undesired final behaviors, such as divergence of opinions or convergence to a neutral opinion.}
\end{remark}
\section{Structural transitions between collective outcomes}
\label{sec:Structural transitions between collective outcomes}
The classification discussed in Section~\ref{sec:Collective decision outcomes in bi-partitioned signed networks} characterizes the limiting outcomes for a given BSN. This section examines how this outcome changes when the strengths of the signed interactions are modified. Such modifications alter the spectrum of \(L_g\), and can, therefore, transition between consensus, clustering, and polarization regimes.

\textcolor{black}{Starting from a BSN in the consensus-compatible regime, increasing the magnitude of antagonistic interactions or weakening selected cooperative interactions alters the spectral properties of \(L_g\). The second smallest eigenvalue approaches zero. At zero, the limiting behavior changes from consensus to clustering. With further increase in the magnitude of inter-subset antagonistic interactions, the network can enter a polarization-compatible regime. Thus, qualitative changes in the collective behavior arise from quantitative changes in the signed edge weights.}

Effective resistance provides an edge-level prediction of these transitions. For a type-1 BSN with \(L_g\succeq0\) and a simple zero eigenvalue, modifying a negative edge with large effective resistance has the \textcolor{black}{most} effect on the spectral margin. 
The following proposition gives the critical value for these edge modifications.
\begin{proposition}
\label{lemma:single_edge_modification}
Let a type-1 BSN satisfy \(L_g \succeq 0\) and
\(\ker L_g=\operatorname{span}\{\mathbf{1}_n\}\). Suppose \(e_r\) is a
negative edge with incidence vector \(b_r\) and maximum effective resistance $r_{\max}=b_r^{T}L_g^{\dagger}b_r$.
Increasing the magnitude of \(e_r\) by
\(\delta^{\star}=1/r_{\max}\) drives the second-smallest eigenvalue of
\(L_g\) to zero, provided the modified network remains type-1 for \(0\leq\delta\leq\delta^{\star}\), \textcolor{black}{where $\delta $ is the scalar edge perturbation parameter for $e_r$}.
\end{proposition}

\begin{proof}
After increasing the magnitude of \(e_r\) by \(\delta>0\), define,
\[
L_g^{\text{pert}}(\delta)
=
L_g-\delta b_rb_r^{T}.
\]

Since \(L_g\succeq0\) has a simple zero eigenvalue, it is positive definite on
\(\mathbf{1}_n^\perp\). 
For any \(\mathbf{y}\perp\mathbf{1}_n\),
Cauchy--Schwarz yields,
\[
(b_r^{T}\mathbf{y})^2
\leq
\left(b_r^{T}L_g^{\dagger}b_r\right)
\mathbf{y}^{T}L_g\mathbf{y}
=
r_{\max}\mathbf{y}^{T}L_g\mathbf{y}.
\]
Hence,
\[
\mathbf{y}^{T}L_g^{\text{pert}}(\delta)\mathbf{y}
\geq
(1-\delta r_{\max})\mathbf{y}^{T}L_g\mathbf{y}.
\]
Therefore, \(L_g^{\text{pert}}(\delta)\) is positive definite on
\(\mathbf{1}_n^\perp\) for \(\delta<1/r_{\max}\).

At \(\delta^{\star}=1/r_{\max}\), let
\(\mathbf{y}^{\star}=L_g^{\dagger}b_r\). Since
\(b_r\perp\mathbf{1}_n\), \(\mathbf{y}^{\star}\neq\mathbf{0}_n\) and,
\[
L_g(\delta^{\star})\mathbf{y}^{\star}
=
b_r-\frac{1}{r_{\max}}b_r
\left(b_r^{T}L_g^{\dagger}b_r\right)
=\mathbf{0}_n.
\]
Moreover, $\mathbf{y}^{\star}\perp\mathbf{1}_n$. 
Thus,
\(L_g(\delta^{\star})\succeq0\) has at least two independent
eigenvectors, and $\lambda_2\!\left(L_g(\delta^{\star})\right)=0.$
\end{proof}
At this threshold, \(L_g\) has an additional zero eigenvalue and the limiting behavior changes from consensus to clustering. Thus, edges with larger effective resistance require smaller modifications to change the collective outcome.
Similarly, the limiting behavior can be varied from polarization to clustering by analyzing the effective resistances in the z-domain.
\vspace{-2mm}
\section{Numerical and empirical validation}
\label{sec:Numerical and empirical validation}
We validate the proposed framework on synthetic and empirical BSNs: stochastic block model networks, a judicial voting network, a U.S. senate bill co-sponsorship network, and an edge-level structural-transition example. 

\begin{figure*}[!htbp]
    \centering
    \begin{subfigure}[b]{5.2cm}
        \centering
        \vspace{-2mm} 
        \includegraphics[width=\linewidth]{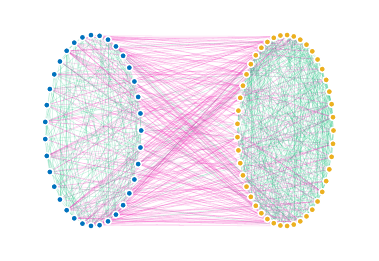}
        \vspace{-7mm} 
        \caption{$\mathcal{G}_4$}
        \label{fig:synthetic_graph_1}
    \end{subfigure}
    \begin{subfigure}[b]{5.2cm} 
        \centering
        \vspace{-2mm} 
        \includegraphics[width=\linewidth]{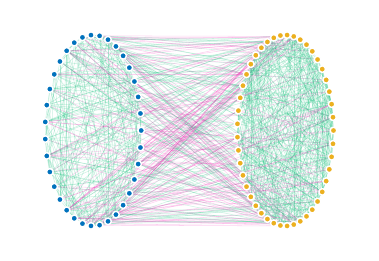}
        \vspace{-7mm} 
        \caption{$\mathcal{G}_5$}
        \label{fig:synthetic_graph_2}
    \end{subfigure}
    \begin{subfigure}[b]{5.2cm}
        \centering
        \vspace{-2mm} 
        \includegraphics[width=\linewidth]{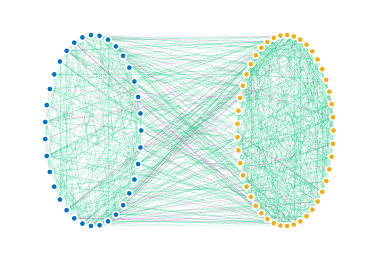}
        \vspace{-7mm} 
        \caption{$\mathcal{G}_6$}
        \label{fig:synthetic_graph_3}
    \end{subfigure}
    \\
    \vspace{2mm}
    \begin{subfigure}[b]{5.2cm}
        \centering
        \vspace{-2mm} 
        \includegraphics[width=\linewidth]{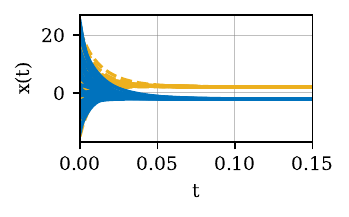}
        \vspace{-7mm} 
        \caption{Polarization of opinions}
        \label{fig:Example_2_solution_plot_polarization}
    \end{subfigure}
    \begin{subfigure}[b]{5.2cm}
        \centering
        \vspace{-2mm} 
        \includegraphics[width=\linewidth]{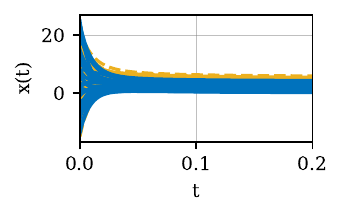}
        \vspace{-7mm} 
        \caption{Clustering of opinions}
        \label{fig:Example_2_solution_plot_clustering}
    \end{subfigure}
    \begin{subfigure}[b]{5.2cm}
        \centering
        \vspace{-2mm} 
        \includegraphics[width=\linewidth]{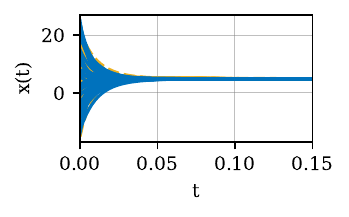}
        \vspace{-7mm} 
        \caption{Consensus of opinions}
        \label{fig:Example_2_solution_plot_consensus}
    \end{subfigure}    
    \vspace{-2mm}
    \caption{SBM random graphs (a), (b), and (c), with inter-subset negative interaction probabilities of 0.9, 0.5, and 0.1, respectively, and their corresponding steady-state behaviors: (d) polarization, (e) clustering, and (f) consensus. In networks $\mathcal{G}_{4}$, $\mathcal{G}_{5}$ and $\mathcal{G}_{6}$, the cooperative and antagonistic interactions are represented using green and pink color, respectively. In these networks, the nodes from subsets $\mathcal{V}_{1}$ and $\mathcal{V}_{2}$ are represented using blue and yellow  colors, respectively.}
    \label{fig:Example_2_plots_sbm}
\end{figure*}

\subsection{Validation using stochastic block model networks}

We first consider synthetic signed networks generated from a two-community stochastic block model. The two communities have sizes \(45\) and \(35\). The probability of an intra-subset edge is \(0.3\), and the probability of an inter-subset edge is \(0.2\). Intra-subset edges are cooperative with weight \(+10\). Inter-subset edges have magnitude \(5\): cooperative inter-subset edges have weight \(+5\), and antagonistic inter-subset edges have weight \(-5\). We vary the probability \(p\) that an inter-subset edge is antagonistic and study the resulting steady-state opinions under the generalized Laplacian flow.
This single generative model produces all three predicted regimes. For \(p=0.9\), the network \(\mathcal G_4\), shown in Figure~\ref{fig:synthetic_graph_1}, is type-2 and the transformed effective resistance matrix \(\Gamma_{\mathcal F_{4z-}}\) is positive definite. Hence, for a large time-scale parameter \(\tau=10^{10}\), the final opinions polarize: individuals in the two subsets converge to opposite polarities with magnitude \(2.1\), as shown in Figure~\ref{fig:Example_2_solution_plot_polarization}. For \(p=0.5\), the network \(\mathcal G_5\), shown in Figure~\ref{fig:synthetic_graph_2}, is type-3. For \(\tau=0.1\), the final opinions form clusters, as shown in Figure~\ref{fig:Example_2_solution_plot_clustering}. For \(p=0.1\), the network \(\mathcal G_6\), shown in Figure~\ref{fig:synthetic_graph_3}, is type-1 and \(\Gamma_{\mathcal F_{6-}}\) is positive definite. Hence, for \(\tau=10^{10}\), the final opinions converge to average consensus at \(5\), as shown in Figure~\ref{fig:Example_2_solution_plot_consensus}. Thus, as the probability of antagonistic inter-subset interactions increases, the steady-state behavior transitions from consensus to clustering and then to polarization.

\begin{figure*}[ht!]
    \centering
    \begin{subfigure}[b]{5.3cm} 
        \centering
        \vspace{-2mm} 
        \includegraphics[width=\linewidth]{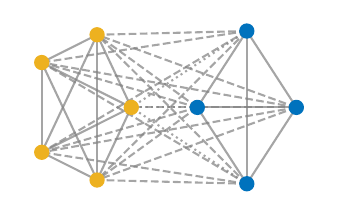}
        \vspace{-7mm} 
        \caption{$\mathcal{G}_7$}
        \label{fig:real_world_graph_1}
    \end{subfigure}
    \begin{subfigure}[b]{5.1cm} 
        \centering
        \vspace{-2mm} 
        \includegraphics[width=\linewidth]{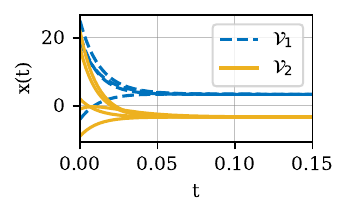}
        \vspace{-7mm} 
        \caption{Polarization of opinions}
        \label{fig:Polarization of opinions}
    \end{subfigure}
    \begin{subfigure}[b]{5.1cm} 
        \centering
        \vspace{-2mm} 
        \includegraphics[width=\linewidth]{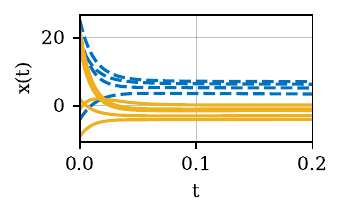}
        \vspace{-7mm} 
        \caption{Clustering of opinions}
        \label{fig:Clustering of opinions-1}
    \end{subfigure}
    \\
    \vspace{2mm}
    \begin{subfigure}[b]{5.3cm} 
        \centering
        \vspace{-2mm} 
        \includegraphics[width=\linewidth]{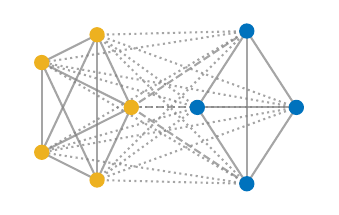}
        \vspace{-7mm} 
        \caption{$\mathcal{G}_8$}
        \label{fig:real_world_graph_cooperative}
    \end{subfigure}
    \begin{subfigure}[b]{5.1cm} 
        \centering
        \vspace{-2mm} 
        \includegraphics[width=\linewidth]{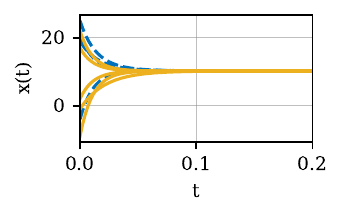}
        \vspace{-7mm} 
        \caption{Consensus of opinions}
        \label{fig:Consensus of opinions}
    \end{subfigure}
    \begin{subfigure}[b]{5.1cm} 
        \centering
        \vspace{-2mm} 
        \includegraphics[width=\linewidth]{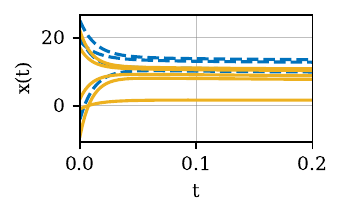}
        \vspace{-7mm} 
        \caption{Clustering of opinions}
        \label{fig:Clustering of opinions-4}
    \end{subfigure}
    \vspace{-2mm}
    \caption{Opinion evolution in the U.S. Supreme Court justices network $\mathcal{G}_7$ in Figure~\ref{fig:real_world_graph_1} and modified U.S. Supreme Court justices network $\mathcal{G}_8$ in Figure~\ref{fig:real_world_graph_cooperative} for varying inter-subset interactions using generalized Laplacian flow. In networks $\mathcal{G}_7$ and $\mathcal{G}_8$, the intra-subset cooperative, inter-subset antagonistic, and inter-subset cooperative interactions are represented with solid, dashed, and dotted edges, respectively. 
    There are no intra-subset antagonistic interactions in these networks. 
    Figures~\ref{fig:Polarization of opinions}- \ref{fig:Clustering of opinions-1} (Figures~\ref{fig:Consensus of opinions}-\ref{fig:Clustering of opinions-4}) show different steady state behaviors for network $\mathcal{G}_{7}$ ($\mathcal{G}_{8}$).
    }
    \label{fig:Example_1_plots_antagonistic}
\end{figure*}
\subsection{Validation using judicial voting network}
We next consider the connected BSN \(\mathcal G_7\), shown in Figure~\ref{fig:real_world_graph_1}, representing voting behavior among nine U.S. Supreme Court justices during 2006--2007 \cite{su2017algorithm,doreian2009partitioning}. Nodes represent justices, positive edges represent frequent voting agreement, and negative edges represent frequent voting disagreement. The prescribed bipartition is indicated by the two node colors in Figure~\ref{fig:real_world_graph_1}.
For \(\mathcal G_7\), we fix intra-subset cooperative weights at \(10\) and inter-subset antagonistic weights at \(-10\). We vary the strength of inter-subset cooperative interactions. When the inter-subset cooperative weights are weak, with value \(1\), the network is type-2 and the transformed effective resistance matrix \(\Gamma_{\mathcal F_{7z-}}\) is positive definite. Hence, for a large time-scale parameter \(\tau=10^{10}\), the final opinions polarize across the prescribed bipartition: the two subsets converge to opposite polarities with magnitude \(3.3\), as shown in Figure~\ref{fig:Polarization of opinions}. When the inter-subset cooperative weights are increased to \(10\), the network becomes type-3. For \(\tau=0.1\), the final opinions form clusters, as shown in Figure~\ref{fig:Clustering of opinions-1}.

Now, we consider the BSN \(\mathcal G_8\), shown in Figure~\ref{fig:real_world_graph_cooperative}, obtained by reversing the polarities of the inter-subset interactions in \(\mathcal G_7\). The prescribed bipartition is kept the same. Whereas inter-subset antagonistic interactions dominate in \(\mathcal G_7\), inter-subset cooperative interactions dominate in \(\mathcal G_8\). For \(\mathcal G_8\), we fix intra-subset cooperative weights and inter-subset cooperative weights at \(+10\), and vary the strength of inter-subset antagonistic interactions. When the inter-subset antagonistic weights are weak, with value \(-1\), the network is type-1 and \(\Gamma_{\mathcal F_{8-}}\) is positive definite. Hence, for \(\tau=10^{10}\), the final opinions converge to consensus at \(10.2\), as shown in Figure~\ref{fig:Consensus of opinions}. When the inter-subset antagonistic weights are increased in magnitude to \(-10\), the network becomes type-3. For \(\tau=0.1\), the final opinions form clusters, as shown in Figure~\ref{fig:Clustering of opinions-4}.

\begin{figure*}[ht!]
    \centering
    \begin{subfigure}[b]{8.7cm} 
        \centering
        \vspace{-2mm} 
        \includegraphics[width=\linewidth]{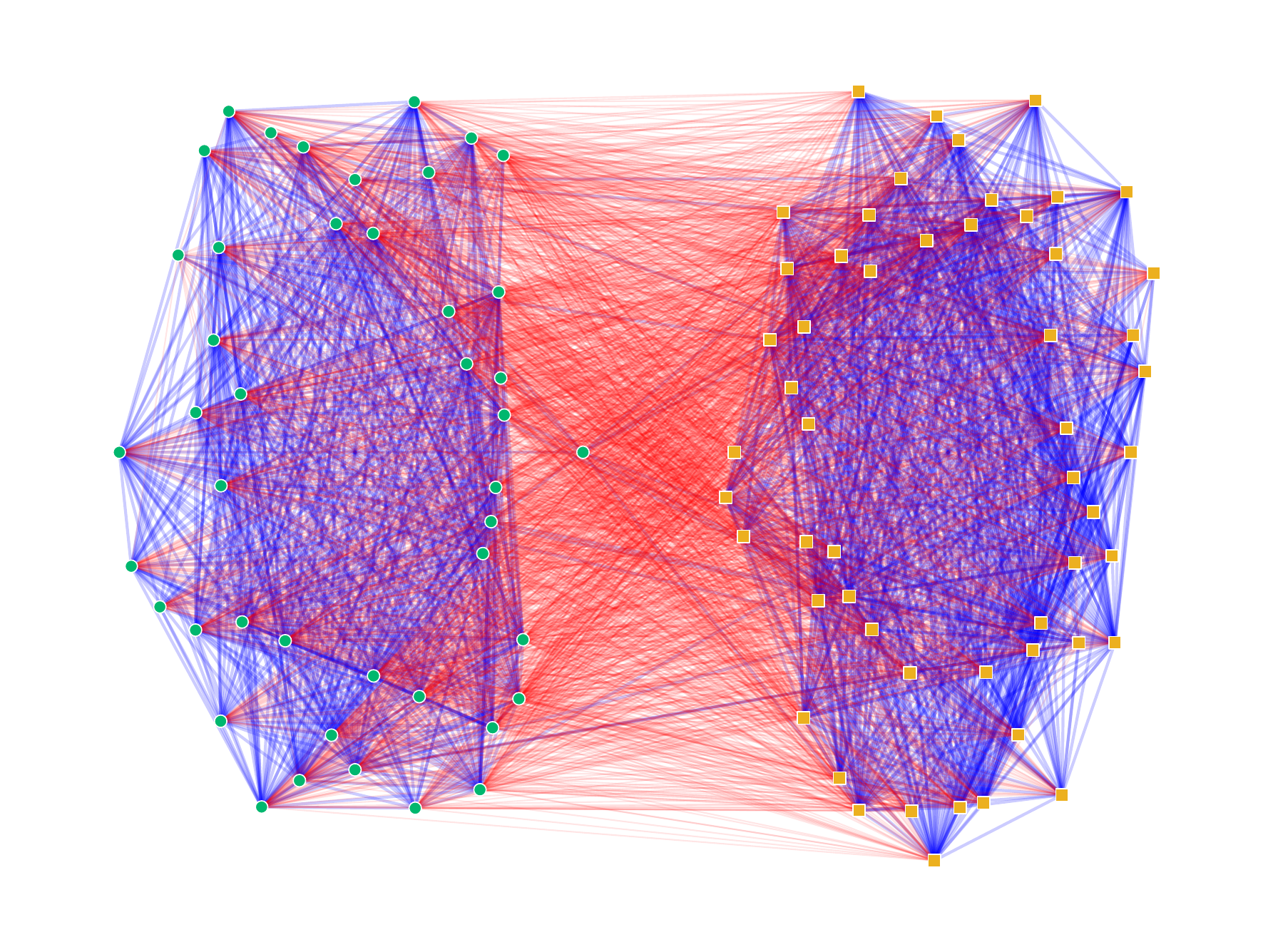}
        \vspace{-7mm} 
        \caption{$\mathcal{G}_{9}$}
        \label{fig:Example_4_Polariation_graph}
    \end{subfigure}
    \begin{subfigure}[b]{8.7cm} 
        \centering
        \vspace{-2mm} 
        \includegraphics[width=\linewidth]{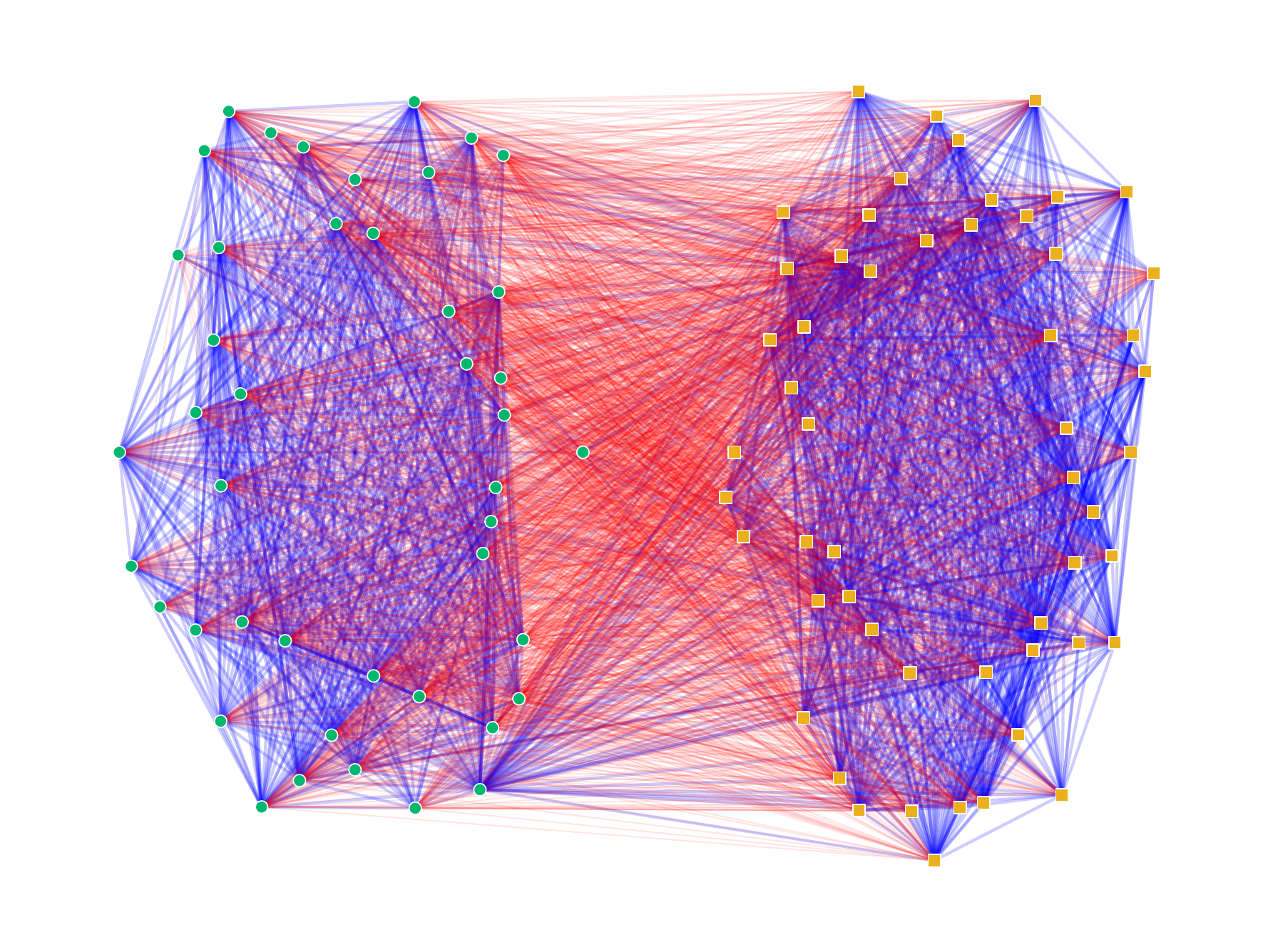}
        \vspace{-7mm} 
        \caption{$\mathcal{G}_{10}$}
        \label{fig:Example_4_Consensus_graph}
    \end{subfigure}
    \\
    \vspace{2mm}
    \begin{subfigure}[b]{5.8cm} 
        \centering
        \vspace{-2mm} 
        \includegraphics[width=\linewidth]{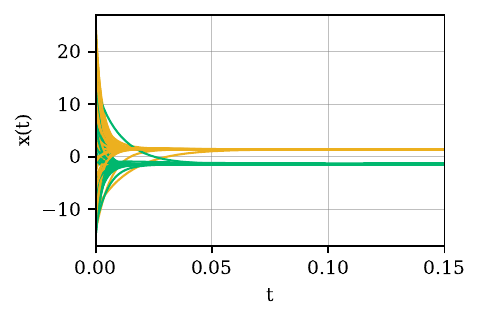}
        \vspace{-7mm} 
        \caption{Polarization of opinions}
        \label{fig:Polarization of opinions-Example_4}
    \end{subfigure}
    \begin{subfigure}[b]{5.8cm} 
        \centering
        \vspace{-2mm} 
        \includegraphics[width=\linewidth]{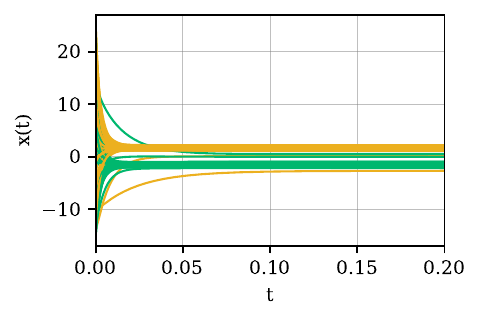}
        \vspace{-7mm} 
        \caption{Clustering of opinions}
        \label{fig:Clustering of opinions-Example4}
    \end{subfigure}
    \begin{subfigure}[b]{5.8cm} 
        \centering
        \vspace{-2mm} 
        \includegraphics[width=\linewidth]{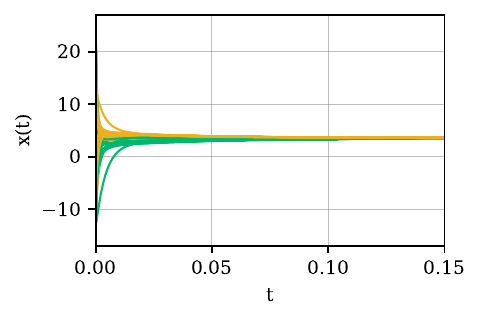}
        \vspace{-7mm} 
        \caption{Consensus of opinions}
        \label{fig:Consensus of opinions-Example4}
    \end{subfigure}
    \vspace{-2mm}
    \caption{Opinion evolution in the network $\mathcal{G}_{9}$ (a) depicting bill co-sponsorship relations in the U.S. senate for the 114th Congress (2015–17, Obama administration). Positive interactions (blue) link senators who tended to promote the same bills together more often than by chance, and negative interactions (red) correspond to those who collaborated less often than at random. Network $\mathcal{G}_{10}$ (b) is made by modifying the inter-subset interactions in $\mathcal{G}_{9}$. The opinion evolution of the individuals in the network $\mathcal{G}_{9}$ can lead to polarization or clustering depending on the interaction weights. The opinion evolution of the individuals in the network $\mathcal{G}_{10}$ leads to consensus. The specific interaction weights of these networks are discussed in the numerical and empirical validation section.}
    \label{fig:Example_4_plots}
\end{figure*}

\subsection{Validation using U.S. senate bill co-sponsorship network}

We also analyze the connected BSN \(\mathcal G_9\), shown in Figure~\ref{fig:Example_4_Polariation_graph}, constructed from bill co-sponsorship relations in the U.S. senate during the \(114^{\text{th}}\) Congress (2015--2017) \cite{talaga2023polarization,neal2020sign}. This session exhibits a strong Democratic--Republican bipartition, with interactions mostly cooperative within parties and mostly antagonistic across parties. Nodes represent senators. Positive edges represent a tendency to co-sponsor the same bills more often than expected, while negative edges represent a tendency to co-sponsor the same bills less often than expected. We use the Democratic--Republican partition, shown in blue and red, as the prescribed bipartition.

For \(\mathcal G_9\), we vary the signed interaction weights and study the resulting steady-state opinions under the generalized Laplacian flow. First, we assign magnitude \(5\) to inter-subset antagonistic and intra-subset cooperative interactions, and magnitude \(1\) to inter-subset cooperative and intra-subset antagonistic interactions. Under these weights, \(\mathcal G_9\) is a type-2 BSN and the transformed effective resistance matrix \(\Gamma_{\mathcal F_{9z-}}\) is positive definite. Hence, for a large time-scale parameter \(\tau=10^{10}\), the final opinions polarize across the Democratic--Republican bipartition, with both subsets reaching opinion magnitude \(1.42\), as shown in Figure~\ref{fig:Polarization of opinions-Example_4}. When all interaction weights are instead assigned magnitude \(5\), the same network becomes type-3. For \(\tau=0.1\), the final opinions form clusters, as shown in Figure~\ref{fig:Clustering of opinions-Example4}.

The original network \(\mathcal G_9\) contains \(2113\) inter-subset interactions, of which only \(18\) are cooperative. Moreover, \(98\) nodes are incident to inter-subset interactions. Therefore, \(\mathcal G_9\) cannot be made type-1 by changing interaction magnitudes alone, because many nodes have only antagonistic inter-subset ties and hence cannot satisfy the type-1 condition \(\alpha_i\ge0\). To construct a consensus-compatible network, we modify \(\mathcal G_9\) by flipping the polarities of \(79\) inter-subset antagonistic interactions so that the resulting network satisfies the type-1 condition at every node. This produces the modified BSN \(\mathcal G_{10}\), shown in Figure~\ref{fig:Example_4_Consensus_graph}.

For \(\mathcal G_{10}\), we assign magnitude \(60\) to cooperative interactions and magnitude \(1\) to antagonistic interactions. Under these weights, \(\mathcal G_{10}\) is a type-1 BSN and \(\Gamma_{\mathcal F_{10-}}\) is positive definite. Hence, for \(\tau=10^{10}\), the final opinions converge to consensus, as shown in Figure~\ref{fig:Consensus of opinions-Example4}. Thus, the senate co-sponsorship network confirms the theoretical predictions on a real-world signed network: dominant inter-subset antagonism produces polarization, balanced signed interactions produce clustering, and sufficiently dominant cooperative connectivity can produce consensus.
\subsection{Validation of structural transition predictions}
\begin{figure}[ht!]
    \centering
    \vspace{-5mm} 
    \includegraphics[width=0.25\linewidth]{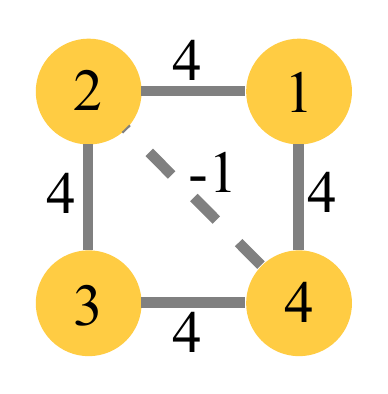}
    \vspace{-4mm}
    \caption{BSN with the intra-subset cooperative, inter-subset cooperative, and inter-subset antagonistic interactions are denoted by solid, dotted, and dashed edges, respectively.} \label{fig:ssb_graph_8}
\end{figure}

We validate the edges transition prediction on the type-1 BSN shown in Figure~\ref{fig:ssb_graph_8}. For this network, \(-L_g\) is positive semidefinite with a simple zero eigenvalue, so the generalized Laplacian flow reaches consensus. The antagonistic interaction \(e_r\) has effective resistance \(r_{\max}=0.33\). Increasing the magnitude of this negative interaction to the threshold value $4$, which is computed by $(1+1/R_{max})$, drives the second eigenvalue to zero, yielding $\sigma(L_g)=\{0,0,8,16\}$. 
The additional zero eigenvalue changes the limiting behavior from consensus to clustering. This confirms that effective resistance identifies which edge modifications can induce structural transitions between collective decision regimes.
\vspace{-4mm}
\section{Discussion and Future Work}
\label{sec:Discussion and Future Works}
This work develops a generalized Laplacian-flow model for collective decision-making in bi-partitioned signed networks. The model captures consensus, polarization, and clustering as distinct steady-state outcomes of a single dynamical system. The analysis shows that these outcomes are determined by the signed interaction structure encoded in the generalized Laplacian \(L_g\), together with effective resistance conditions that characterize when the relevant consensus- or polarization-compatible modes are present.

A key implication is that antagonistic interactions do not have a single universal effect. Depending on their placement and strength relative to cooperative interactions, they can support polarization, produce clustered opinion profiles, or be absorbed within a consensus-compatible structure. Effective resistance provides an edge-level way to quantify this role: it identifies how antagonistic interactions affect the spectral margins that separate consensus, polarization, and clustering regimes. Thus, the theory connects local signed interactions to global collective outcomes.

The results also show that collective behavior can change qualitatively under relatively small structural modifications. Changing the strength or sign of selected interactions can alter the spectrum of \(L_g\), causing a transition from consensus to clustering, from clustering to polarization, or conversely toward consensus. This provides a structural explanation for how gradual changes in social or network interactions can produce abrupt changes in collective behavior.

Several extensions remain open. The present analysis focuses on undirected signed networks with a prescribed bipartition. Future work will extend the framework to directed networks, time-varying interactions, uncertain or learned bipartitions, and optimal intervention problems in which one seeks minimal structural changes to achieve a desired collective decision outcome.


%








\ifCLASSOPTIONcaptionsoff
  \newpage
\fi



\bibliographystyle{IEEEtran}
\bibliography{mybibiliography}
%



%

\newpage

\end{document}